\documentclass[12pt]{article}
\newcommand{\T}{\overline{T}}
\newcommand{\Om}{\overline{\Omega}}
\renewcommand{\d}{{\rm d}}
\newcommand{\ve}{\varepsilon}
\newcommand{\es}{\epsilon(\sigma^2,\sigma^3)}
\newcommand{\Gs}{\Gamma_2(\sigma^2,\sigma^3)}
\newcommand{\barGs}{\overline{\Gamma}_2(\sigma^2,\sigma^3)}
\begin{document}

\title{Area Regge calculus and continuum limit}
\author{V.M.Khatsymovsky \\
 {\em Budker Institute of Nuclear Physics} \\ {\em
 Novosibirsk,
 630090,
 Russia}
\\ {\em E-mail address: khatsym@inp.nsk.su}}
\date{}
\maketitle
\begin{abstract}
Encountered in the literature generalisations of general
relativity to independent area variables are considered,
the discrete (generalised Regge calculus) and continuum
ones. The generalised Regge calculus can be either with
purely area variables or, as we suggest, with area
tensor-connection variables. Just for the latter, in
particular, we prove that in analogy with corresponding
statement in ordinary Regge calculus (by Feinberg,
Friedberg, Lee and Ren), passing to the (appropriately
defined) continuum limit yields the generalised continuum
area tensor-connection general relativity.
\end{abstract}
\newpage
The idea that Regge calculus should be formulated in terms
of the areas of the triangles instead of the edge lengths
\cite{Rov} originates from the attempts to generalise the
3-dimensional Ponzano-Regge model of quantum gravity
\cite{PonReg,TurVir} to the physical 4-dimensional case
\cite{Oog,CraKauYet}. One way of treating the area
variables is by imposing certain geometrical constraints
enforcing them to be still expressible in terms of the edge
lengths \cite{Mak}. Principially new theory arises,
however, if one treats the areas as the fundamental and
independent variables \cite{BarRocWil}. The basis for
possibility and naturalness of such treatment in the
physical 4-dimensional case is simple observation that the
number 10 of the edges of a 4-simplex is the same as the
number of it's triangular faces. Therefore the both sets,
those of area and length variables turn out to be
expressible in terms of each other inside a given
4-simplex, at least locally. The requirement for the
neighbouring 4-simplices to have coinciding lengths of (up
to 6) common edges is relaxed to the requirement of having
only coinciding areas of (up to 4) common triangles
instead. This leaves the lengths ambiguous in general, but
still allows to define the dihedral angles in each the
4-simplex. Therefore the angle defects $\varphi_\Delta$ on
the triangles $\Delta$ are defined as functions of the
triangle areas $A_\Delta$, and Regge calculus action still
can be written out,
\begin{equation}
\label{S-A}                                              %1
S_{\rm area} = \sum_\Delta{A_\Delta\varphi_\Delta(\{
A_\Delta\})}.
\end{equation}

\noindent Variation w.r.t. $A_\Delta$ is performed as in
the ordinary Regge calculus, but since variations $\delta
A_\Delta$ for the different $\Delta$ are now independent
values, the equations of motion give $\varphi_\Delta$ = 0
\cite{BarRocWil}. Because of the lack of metric and of
the usual geometric interpretation of $\varphi_\Delta$,
this does not mean flat spacetime \cite{BarRocWil}.
Moreover, it was mentioned \cite{RegWil} that diagonalising
second variation of the action (\ref{S-A}) as bilinear form
of $\delta A_\Delta$ around a (distorted) hypercubic
lattice results in the same dynamical content (the number
of the dynamical degrees of freedom) as in the usual Regge
calculus based on the edge length variables \cite{RocWil}.

An interesting question is that of possible continuum
counterpart of the area Regge calculus with the action
(\ref{S-A}). On the other hand, the Hilbert-Palatini form
of the usual general relativity admits generalisation to
area variables in a natural way. Indeed, write the action
in the form
\begin{equation}
\label{S-HP}                                             %2
S_{\rm HP}={1\over 4}\int\! d^{4}x\: \pi_{ab}^{\lambda\mu}
[{\cal D}_{\lambda},{\cal D} _{\mu}]^{ab}
\end{equation}

\noindent where ${\cal D}_{\lambda}$ = $\partial_{\lambda}$
+ $\omega_{\lambda}$ (in fundamental representation) is
covariant derivative, and $\omega_{\mu}^{ab}$ = $-\omega
_{\mu}^{ba}$ is element of $so(3,1)$, Lie algebra of
$SO(3,1)$ group in the Lorentzian case or an element of
$so(4)$, Lie algebra of $SO(4)$ in the Euclidean case.
$\lambda$, $\mu$, \ldots = 1, 2, 3, 4 are coordinate
indices and $a$, $b$, \ldots = 1, 2, 3, 4 are local ones.
The antisymmetric in $a$, $b$ and in $\lambda$, $\mu$ {\it
area tensor} $\pi_{ab}^{\lambda\mu}$ is subject to the
tensor relation
\begin{equation}
\label{pi-pi}                                            %3
\pi_{ab}^{\lambda\mu}\pi_{cd}^{\nu\rho}\epsilon^{abcd}
\sim\epsilon^{\lambda\mu\nu\rho}.
\end{equation}

\noindent This equation simply ensures that a tetrad $e_
\lambda^a$ exists so that $\pi_{ab}^{\lambda\mu}$ is a
bivector,
\begin{equation}                                         %4
\pi_{ab}^{\lambda\mu}={1\over 2}\epsilon^{\lambda\mu\nu
\rho}\epsilon_{abcd}e_\nu^ce_\rho^d.
\end{equation}

\noindent More often treated as area tensor is the twice
dual to $\pi_{ab}^{\lambda\mu}$,
\begin{equation}
\label{pi-v}                                             %5
v_{\lambda\mu}^{ab} = \frac{1}{4}\epsilon^{abcd}
\epsilon_{\lambda\mu\nu\rho}\pi_{cd}^{\nu\rho},
\end{equation}

\noindent which in the tetrad formalism reduces to
$e_\lambda^ae_\mu^b$ - $e_\lambda^be_\mu^a$. Generalisation
simply amounts to omitting the eq. (\ref{pi-pi}) so that
the components of area tensor become independent variables.
In fact, this generalisation is the subject of study in the
literature when Ashtekar formalism \cite{Ash} is discussed.
Indeed, Ashtekar formalism can be obtained by separating
self- and antiselfdual (over local indices) parts of
$\omega_\lambda^{ab}$, $\pi_{ab}^{\lambda\mu}$ in eqs.
(\ref{S-HP}, (\ref{pi-pi}), the eq. (\ref{pi-pi}) being an
issue point for the {\it reality conditions} (in the case
of the Lorentzian signature) in this formalism \cite{Pel,
Rov1}. The hope is that this (in fact, more simple)
formalism with unrestricted area tensor can be solved and
the reality conditions can be imposed anyhow afterwards to
select a real section of the complex phase space
\cite{Rom}. An yet unresolved problem encountered in this
way is how do classical configurations of the gravitational
field like gravitons arise. In this respect, there is the
difference from the area Regge calculus based on the action
(\ref{S-A}) where the dynamical degrees of freedom probably
match those in the ordinary length Regge calculus as
mentioned above.

Thus, we have the two generalisations of general relativity
to independent area variables, the discrete (\ref{S-A}) and
continuum (\ref{S-HP}) ones which use the purely area and
area tensor - connection variables, respectively. What is
the connection between these two? When linking the ordinary
length (metric) and tetrad - connection formalisms one uses
the eq. $g_{\lambda\mu}$ = $e_\lambda^ae_\mu^a$ relating
the different sets of independent variables; then it is
noted that the metric tensor $g_{\lambda\mu}$ has the same
number of components as the number of edges of a 4-simplex.
Now in analogy one could like to construct the "area metric
tensor" $h_{\lambda\mu\nu\rho}$ = $v_{\lambda\mu}^{ab}
v_{\nu\rho}^{ab}$ which being viewed as a $6\times 6$
symmetrical matrix w.r.t. antisymmetrical pairs $[\lambda
\mu]$ and $[\nu\rho]$ has 21 independent components, not
the same as the number 10 of independent areas of a
4-simplex. Therefore we cannot say that there is a natural
equivalence between the both considered area variable
formalisms. At least, either the 4-simplex cannot serve as
an elementary cell of the corresponding "area geometry"
(defined by $h_{\lambda\mu\nu\rho}$) or the field
$h_{\lambda\mu\nu\rho}$ is subject to some additional
constraints. Indirectly, inequivalence between the two
formalisms displays also in the above mentioned different
dynamical content of them.

On the other hand, the tetrad-connection (including the
case of self-dual connection) representation of ordinary
Regge calculus has been suggested by the author \cite{Kha},
\begin{equation}
\label{S-VR}                                             %6
S(V,\Omega) = \sum_{\sigma^2}{|V_{\sigma^2}|\arcsin\frac
{V_{\sigma^2}\ast R(\Omega)}{|V_{\sigma^2}|}}
\end{equation}

\noindent where $V_{\sigma^2}^{ab}$ are the bivectors of
the 2-faces $\sigma^2$, $|V|^2$ $\equiv$ ${1\over 2}V^{ab}
V^{ab}$, $R_{\sigma^2}(\Omega)$ is the product of the SO(4)
in the Euclidean (SO(3,1) in the Lorentzian) case matrices
$\Omega_{\sigma^3}$ living on the 3-faces $\sigma^3$ taken
along the loop enclosing the given 2-face $\sigma^2$,
$V\ast R$ $\equiv$ ${1\over 4}V^{ab}R^{cd}\epsilon_{abcd}$.
Strictly speaking, the bivector carries one else subscript
$\sigma^4$ as $V_{\sigma^2,\sigma^4}$ indicating the local
frame where the bivector of a given 2-face $\sigma^2$ is
defined. Also the eq. (\ref{S-VR}) should be accomplished
with geometrical constraints ensuring the bivector form of
$V_{\sigma^2}^{ab}$. The form of these constraints can be
taken very simple, linear and bilinear, although at the
price of rather large number of them, if we extend the set
of $V_{\sigma^2,\sigma^4}^{ab}$ to all $\sigma^4$
$\supset$ $\sigma^2$ \cite{Kha1}. Generalisation to
independent area tensor formalism is by simply omitting
these constraints, and we are left with eq. (\ref{S-VR})
alone with freely varied $V_{\sigma^2}$ and $\Omega
_{\sigma^3}$ variables.

Write out the table of actions for possible versions of
area general relativity.
%\begin{table}
\begin{flushleft}
\begin{tabular}{|l|c|c|}          \hline
~~~~variab- &      &             \\
disc-~~~~les& area & area tensor \\
reteness  &      & -connection \\ \hline
          &      &             \\
continuum &  ?   & $\int\pi(\partial\omega+\omega\wedge
\omega)$ \\
          &      &                            \\ \hline
          &      &              \\
discrete  & $\sum A\varphi(A)$ & $\sum |V|\arcsin\frac{
V\ast R}{|V|}$ \\
          &      &               \\ \hline
\end{tabular}
\unitlength 1mm
\begin{picture}(100,50)
\put(0,97.5){\line(3,-2){23.5}}
\end{picture}
\end{flushleft}
%\end{table}
\vspace{-50mm}
An interesting problem besides that of filling in the upper
left cell of the table is that of establishing
correspondence between the different cells. In the present
paper we obtain the area tensor-connection continuum action
(\ref{S-HP}) from the discrete one (\ref{S-VR}) in the
(properly defined) continuum limit (an analog of the
theorem by Feinberg, Friedberg, Lee and Ren
\cite{FeiFriLeeRen} for the ordinary edge length Regge
calculus).

First choose Regge lattice of a certain periodic structure
used in \cite{RocWil}. Topologically, the Regge manifold
periodic cell is a 4-cube divided into 24 4-simplices
sharing the hyperbody diagonal. Let the indices $\lambda$,
$\mu$, \ldots = 1, 2, 3, 4 label the cube edges emerging
from a vertex $O$ along the corresponding coordinate axes.
The $T_\lambda$, $T_\lambda^{-1}$ = $\T_\lambda$
are operators of the translations to the two neighbouring
vertices in the positive and negative directions of
$\lambda$. Introduce multiindices $A$, $B$, $C$, \ldots ,
the unordered sequences of different indices, e. g. $A$ =
$(\lambda\mu\ldots\nu)$. The link (1-simplex) connecting
the points $O$ and $T_\lambda T_\mu\ldots T_\nu O$ will be
labelled just by $A$ while the $k$-simplex at $k$ $>$ 1
spanned by the links $A_1$, $(A_1A_2)$, \ldots , $(A_1A_2
\ldots A_k)$ will be denoted by the ordered sequence of
multiindices $[A_1A_2\ldots A_k]$. Here the symbol $(A_1A_2
\ldots A_i)$ means multiindex composed of all the indices
encountered in $A_1$, $A_2$, \ldots , $A_i$. If there can
be no confusion, the round and square brackets will be
omitted: notation '$[AB]$' is equivalent to 'the 2-simplex
$AB$' etc. On the whole, we have the following simplices
attributed to the given vertex $O$:\\
(i) 15 links $\lambda$, $\lambda\mu$, $\lambda\mu\nu$,
1234;\\
(ii) 50 2-simplices (triangles) $\lambda\mu$, $(\lambda\mu)
\nu$, $\lambda(\mu\nu)$, $\lambda(\mu\nu\rho)$, $(\lambda
\mu)(\nu\rho)$, $(\lambda\mu\nu)\rho$;\\
(iii) 60 3-simplices $\lambda\mu\nu$, $(\lambda\mu)\nu
\rho$, $\lambda(\mu\nu)\rho$, $\lambda\mu(\nu\rho)$ (the
latter three symbols will be also more briefly written as
$\d\nu\rho$, $\lambda\d\rho$, $\lambda\mu\d$,
respectively, the "d" meaning "diagonal");\\
(iv) 24 4-simplices $\lambda\mu\nu\rho$.

To each oriented 3-simplex $\sigma^3$ shared by the
4-simplices $\sigma_1^4$ and $\sigma_2^4$ we assign the
SO(4) (SO(3,1)) matrix $\Omega_{\sigma^3}^{\epsilon(\sigma
_1^4,\sigma_2^4)}$ in the Euclidean (Lorentzian) case which
acts from the local frame of $\sigma_1^4$ to that of
$\sigma_2^4$; the choice of $\epsilon(\sigma_1^4,
\sigma_2^4)$ = $\pm 1$ just specifies orientation of
$\sigma^3$ = $\sigma_1^4\cap\sigma_2^4$. To each 2-simplex
$\sigma^2$ we assign a simplex $\sigma^4$ $\supset$
$\sigma^2$ in the frame of which area tensor $V_{\sigma^2}$
is defined; then $R_{\sigma^2}$, the product of matrices
$\Omega_{\sigma^3}^{\pm 1}$ acting along the loop enclosing
$\sigma^2$ acts from this $\sigma^4$ to itself. Our choice
of orientation of the 3-simplices and of the frames of
definition of area tensors corresponds to the following
expressions for the curvature matrices,
\begin{eqnarray}
\label{R}                                                %7
R_{41} & = & \Om_{413}(\T_2\Om_{241})(\T_{23}\Om_{\d41})
(\T_3\Omega_{341})\Omega_{412}\Omega_{41\d},\nonumber\\
R_{4(23)} & = & \Om_{4\d1}\Om_{423}(\T_1\Omega_{14\d})
\Omega_{432},\nonumber\\
R_{23} & = & \Om_{23\d}\Om_{231}(\T_4\Om_{423})(\T_{14}
\Omega_{\d23})(\T_1\Omega_{123})\Omega_{234},\nonumber\\
R_{2(43)} & = & \Om_{2\d1}\Om_{234}(\T_1\Omega_{12\d})
\Omega_{243},\nonumber\\
R_{(24)3} & = & \Om_{\d31}\Om_{243}(\T_1\Omega_{1\d3})
\Omega_{423},\\
R_{1(32)} & = & \Om_{132}(\T_4\Om_{41\d})\Omega_{123}
\Omega_{1\d4},\nonumber\\
R_{1(432)} & = & \Om_{1\d4}\Omega_{12\d}\Omega_{1\d3}
\Omega_{14\d}\Om_{1\d2}\Om_{13\d},\nonumber\\
R_{(14)(32)} & = & \Om_{14\d}\Omega_{\d23}\Omega_{41\d}
\Om_{\d32},\nonumber\\
\lefteqn{\rm\ldots~cycle(1,2,3)~\ldots,}
\nonumber\\
R_{4(123)} & = & \Omega_{4\d3}\Om_{42\d}\Omega_{4\d1}
\Om_{43\d}\Omega_{4\d2}\Om_{41\d}.\nonumber
\end{eqnarray}

\noindent These eqs. define 25 expressions. The remaining
half of the whole number 50 of curvature matrices can be
obtained by permuting groups of indices: if $R_{(\lambda
\ldots\mu)(\nu\ldots\rho)}$ = $\prod$ $T_{(\ldots)}^{\pm
1}$ $\Omega_{\ldots\lambda\ldots\mu\nu\ldots\rho\ldots}
^{\pm 1}$ then $\overline{R}_{(\nu\ldots\rho)(\lambda\ldots
\mu)}$ = $\prod$ $T_{(\ldots)}^{\pm 1}$ $\Omega_{\ldots\nu
\ldots\rho\lambda\ldots\mu\ldots}^{\pm 1}$. This completely
defines the tensor area-connection Regge-type action $S(V,
\Omega)$.

The variables $\Omega_{\sigma^3}$, $V_{\sigma^2}$ are in
the natural way the functions of the vertices $O$ to which
the given $\sigma^3$, $\sigma^2$ are attributed. To pass to
the continuum limit we suppose that these variables are the
particular values of some smooth functions $\Omega(x)$,
$V(x)$ on the spacetime continuum taken at the locations of
the vertices. Then we uniformly tend the coordinate
differences between the neighbouring vertices to zero (thus
enlarging the number of vertices in any finite region). The
analog of the derivative, $T_\lambda$ $-$ 1 should be of
the order of $\ve$, the typical coordinate difference
between the neighbouring vertices. For the continuum limit
leading to an expression of the type of $S_{\rm HP}$ be
defined it turns out natural to ascribe the following
orders of magnitude in $\ve$ $\rightarrow$ 0 to the
discrete values in question,
\begin{equation}
\label{V-Om-eps}                                         %8
V = O(\ve^2),~w = O(\ve)~(\exp{w}\equiv\Omega),
\end{equation}
\begin{equation}
\label{V}                                                %9
V_{BA} = -V_{AB}+O(\ve^3),
\end{equation}
\begin{equation}
\label{Omega}                                           %10
w_{ABC} = w_{BAC}+O(\ve^2) = w_{ACB}+O(\ve^2),
\end{equation}

\noindent which correspond to the naive considerations that
we deal with tensors $V$ of the closely located almost
parallel (up to $O(\ve)$) 2-simplices $AB$, $BA$ and
matrices $w$ for the parallel vector transport in the
almost parallel directions orthogonal to the 3-simplices
$ABC$, $BAC$, $ACB$ at almost equal (up to $O(\ve^2)$)
distances $O(\ve)$ separating centers of almost similar
4-simplices. Of course, a geometric interpretation is valid
only for the particular case when area variables correspond
to certain edge length (tetrad) ones, but the orders of
magnitude presented turn out to be justified in what
follows from the purely formal computational grounds as
well.

In the continuum limit the area tensors $V_{\lambda\mu}$
directly correspond to $v_{\lambda\mu}$ in $S_{\rm HP}$,
eqs. (\ref{S-HP}), (\ref{pi-v}) (more exactly, to $\ve^2
v_{\lambda\mu}$). In order to reduce area tensors of other
2-simplices to $v_{\lambda\mu}$ we need relations of the
type $V_{\lambda(\mu\nu)}$ = $V_{\lambda\mu}$ $+$
$V_{\lambda\nu}$. The latter in the usual tetrad formalism
would follow (in the leading order in $\ve$ when one can
neglect rotations needed to express different tensors in
the same frame) from representation of the link vector
$l_{\mu\nu}^a$ as the sum of $l_\mu^a$ and $l_\nu^a$. What
consequence of the theory could, in principle, provide us
with relations of such type is only the Gauss law
accessible under some assumptions via eqs. of motion for
$\Omega$ from $S(V,\Omega)$. That is, the limiting
expression of the type of $S_{\rm HP}$ can be obtained only
upon partial use of the eqs. of motion. For that we take
first and second orders in the expansion of $S(V,\Omega)$
in $w$,
\begin{equation}
\label{S1+S2}                                           %11
S = S^{(1)}(V,w) + S^{(2)}(V,w,w)
\end{equation}

\noindent where $S^{(1)}$, $S^{(2)}$ are first and second
order forms in $w$, both linear in $V$. According to eq.
(\ref{V-Om-eps}) higher orders in $w$ give vanishing at
$\ve$ $\rightarrow$ 0 contribution as a sum of the terms
$O(\ve^5)$ over cells number of which in a fixed finite
region is $O(\ve^{-4})$. Partially using eqs. of motion for
$w$ reduces this equation to
\begin{equation}
\label{S2}                                              %12
-S = +S^{(2)}(V,w,w).
\end{equation}

\noindent Here it is sufficient to know $V$ in the leading
order in $\ve$. Write out the eq. of motion for $\Omega$
which plays the role of the Gauss law. For any $\sigma^2$
and $\sigma^3$ $\supset$ $\sigma^2$ we have $R_{\sigma^2}$
= $(\Gamma_1\Omega_{\sigma^3}\Gamma_2)^\epsilon$ where
$\Gamma_1$, $\Gamma_2$ $\in$ SO(4) (SO(3,1)) and $\epsilon$
= $\pm 1$ are functions of $\sigma^2$, $\sigma^3$. Then for
a given $\sigma^3$
\begin{eqnarray}                                        %13
\lefteqn{\hspace{-20mm}\sum_{\sigma^2\subset\sigma^3}\es\Gs
\left[V(\sigma^2){\rm tr}R(\sigma^2) - V(\sigma^2)R(\sigma
^2)^{-\es}\right.}\nonumber\\
& & \left.\mbox{} - R(\sigma^2)^{\es}V(\sigma^2)\right]
\barGs\frac{1}{\cos\varphi(\sigma^2)} = 0,\\
& & \hspace{-20mm}\sin\varphi(\sigma^2) = \frac{V(\sigma^2)
\ast R(\sigma^2)}{|V(\sigma^2)|}\nonumber
\end{eqnarray}

\noindent \cite{Kha}. With taking into account the eq.
(\ref{Omega}) we have $R(\sigma^2)$ = {\bf 1} + $O(\ve^2)$
(connection matrices enter the products defining curvature
matrices as pairs $\Omega_1$, $\Om_2$ where $\Omega_1$ and
$\Omega_2$ are approximately equal up to {\bf 1} +
$O(\ve)$). Therefore in the leading and next-to-leading
orders in $\ve$ we have the Gauss law
\begin{equation}
\label{Gauss}                                           %14
\sum_{\sigma^2\subset\sigma^3}\es\Gs V(\sigma^2)\barGs =
O(\ve^4).
\end{equation}

\noindent On our particular Regge lattice and with taking
into account (\ref{V}) the leading order reads
\begin{equation}
\label{Gauss0}                                          %15
V_{\lambda(\mu\nu)} + V_{\nu(\lambda\mu)} + V_{\mu\lambda}
+ V_{\mu\nu} = O(\ve^3)
\end{equation}

\noindent and
\begin{equation}
\label{Gauss0'}                                         %16
V_{(\lambda\mu)(\nu\rho)} + V_{\rho(\lambda\mu\nu)} +
V_{\nu(\lambda\mu)} + V_{\nu\rho} = O(\ve^3).
\end{equation}

\noindent Introduce the quantities $\delta_{AB}$ which
correct the naive (inspired by the tetrad formalism)
decompositions of the tensors $V_{AB}$ in terms of those
tensors of the simplest triangles $\lambda\mu$; for
example,
\begin{equation}                                        %17
V_{\lambda(\mu\nu)} = V_{\lambda\mu} + V_{\lambda\nu} +
\delta_{\lambda(\mu\nu)}.
\end{equation}

\noindent Then
\begin{eqnarray}
\label{del}                                             %18
\delta_{\lambda(\mu\nu)}+\delta_{\nu(\lambda\mu)} & = &
O(\ve^3),\\
\label{del'}                                            %19
\delta_{(\lambda\mu)(\nu\rho)} + \delta_{\rho(\lambda\mu
\nu)} + \delta_{\nu(\lambda\mu)} & = & O(\ve^3)
\end{eqnarray}

\noindent from the eqs. (\ref{Gauss0}) and (\ref{Gauss0'}),
respectively. The approximate antisymmetry of $V_{AB}$ w.
r. t. $A$, $B$ (eq. (\ref{V})) also holds for $\delta
_{AB}$. The eq. (\ref{del}) means that $\delta_{\lambda(\mu
\nu)}$ (symmetrical w. r. t. the second and third indices
by definition) is approximately antisymmetrical in the
first and second (third) indices. It is easy to see that
such the quantity can be equal only to zero with the same
accuracy. Substituting this result into the eq.
(\ref{del'}) we see that $\delta_{(\lambda\mu)(\nu\rho)}$
is approximately symmetrical in $\lambda$, $\nu$ and,
consequently, in $\mu$, $\rho$ and, therefore, in the
multiindices $(\lambda\mu)$, $(\nu\rho)$. Together with the
above stated antisymmetry in multiindices this means that
the related $\delta$'s vanish up to $O(\ve^3)$ as well.
Thus, the naive expressions expectable from the analogy
with projecting area bivectors onto the coordinate planes
in the usual tetrad formalism hold in our case too,
\begin{eqnarray}
V_{\lambda(\mu\nu)} & = & V_{\lambda\mu} + V_{\lambda\nu}
+ O(\ve^3),\\                                           %20
V_{\lambda(\mu\nu\rho)} & = & V_{\lambda\mu}+V_{\lambda\nu}
+V_{\lambda\rho} + O(\ve^3),\\                          %21
V_{(\lambda\mu)(\nu\rho)} &=&V_{\lambda\nu}+V_{\lambda\rho}
+V_{\mu\nu}+V_{\mu\rho} + O(\ve^3).                     %22
\end{eqnarray}

\noindent Remarkably is that these more stringent than the
Gauss law relations turn out to be the consequences of it
in some assumptions (that the variables vary slowly between
neighbouring simplices).

Now it is straightforward to substitute the decompositions
of area tensors $V_{AB}$ over "elementary" ones $V_{\lambda
\mu}$ obtained into the second order in $w$ part of the
action, eq. (\ref{S2}). The latter, in turn, is contributed
by the bilinear in $w$ antisymmetric parts of the curvature
matrices $R_{AB}^{(2)}$. For example,
\begin{equation}                                        %23
R_{23}^{(2)} = [w_{23\d}, w_{231}] + [w_{23\d}, w_{423}]
+ [w_{231}, w_{423}].
\end{equation}

\noindent Here translation operators $T_\lambda$ are
substituted by the unity with the accuracy of $O(\ve^3)$.
The proportional to $V_{23}$ bilinear in $w$ parts of the
action are contained in $V_{AB}\ast R_{AB}^{(2)}$ at $AB$
= 23, (21)3, (24)3, 2(31), 2(34), (214)3, 2(143), (21)(34),
(24)(13). Collecting these contributions we find
\begin{equation}
\label{VR}                                              %24
\sum_{AB}V_{AB}\ast R_{AB}^{(2)} = V_{23}\ast [w_1, w_4] +
\ldots = \frac{1}{4}\sum_{\lambda\mu\nu\rho}V_{\lambda\mu}
\ast[w_\nu, w_\rho]\epsilon_{\lambda\mu\nu\rho}
\end{equation}

\noindent where
\begin{eqnarray}                                        %25
w_1 & = &\mbox{} w_{234} - w_{34\d} + w_{24\d} - w_{23\d},
\nonumber\\
\lefteqn{\rm\ldots~cycle(1,2,3)~\ldots,}\\
-w_4 &=&w_{123} + w_{12\d} + w_{23\d} + w_{13\d}.\nonumber
\end{eqnarray}

Thus the continuum action as bilinear form of $\omega$
(following upon partial use of the eqs. of motion) is
obtained from the corresponding discrete version (\ref{S2})
in the continuum limit, by identifying
$V_{\lambda\mu}$ with $\ve^2v_{\lambda\mu}$ and $w_\lambda$
with $\ve\omega_\lambda$. To reproduce the action in the
standard form with independent $\omega$ we need the
equations of motion which should be derived themselves from
the discrete version and then used in backward direction as
compared to how we have obtained the discrete action as
bilinear in $w$, eq. (\ref{S2}) from the original linear
plus bilinear expression with independent $w$, eq.
(\ref{S1+S2}). The equations of motion are just the Gauss
law which has been already written out, eq. (\ref{Gauss}),
but now the linear in $w$ part is of interest, i. e. it
should be expanded to the next-to-leading order in $\ve$.
The Gauss law in the continuum theory expresses closure of
the surface of infinitesimal 3-cube. Consider the 3-cube of
the Regge lattice laying, say, in the 123-hyperplane and
composed of six 3-simplices (123 and permutations) and
write out the eq. (\ref{Gauss}) for each of these
simplices, e. g.
\begin{equation}                                        %26
-\Omega_{12\d}V_{12}\Om_{12\d} + \Omega_{1\d4}V_{1(32)}
\Om_{1\d4} + T_1(\Omega_{234}V_{23}\Om_{234}) -
\Omega_{\d34}V_{(21)3}\Om_{\d34} = O(\ve^4)
\end{equation}

\noindent for $\sigma^3$ = 123 and
\begin{equation}                                        %27
\Omega_{13\d}V_{13}\Om_{13\d} - V_{1(32)} + V_{(13)2} -
T_1(\Omega_{324}V_{32}\Om_{324}) = O(\ve^4)
\end{equation}

\noindent for $\sigma^3$ = 132. Excluding $V_{1(32)}$ from
these equations we get to the linear order in $w$
\begin{eqnarray}                                        %28
-V_{12}+V_{13}+V_{(13)2}-V_{(21)3}+T_1(V_{23}-V_{32})
-[w_{12\d},V_{12}]+[w_{1\d4}+w_{13\d},V_{13}]& &\nonumber\\
+[w_{1\d4},
V_{(13)2}]-[w_{1\d4}+w_{324},V_{32}]+[w_{234},V_{23}]
-[w_{\d34},V_{(21)3}]=O(\ve^4). & &
\end{eqnarray}

\noindent This can be summed up with two other cyclic
permutations of 1, 2, 3. In the terms $[w,V]$ the leading
in $\ve$ accuracy of the definition of $V$ is sufficient,
and these area tensors can be reduced to $V_{\lambda\mu}$.
In the terms of zero order in $w$ the area tensors other
than $V_{\lambda\mu}$ are cancelled in the overall sum.
Thus we obtain
\begin{equation}                                        %29
(T_1-1)V_{23}+[w_1,V_{23}]+{\rm cycle}(1,2,3)=O(\ve^4).
\end{equation}

\noindent This just reduces to the $\rho$ = 4 component of
the continuum Gauss law,
\begin{equation}                                        %30
(\partial_\lambda v_{\mu\nu}+[\omega_\lambda,v_{\mu\nu}])
\epsilon^{\lambda\mu\nu\rho} = 0.
\end{equation}

\noindent This allows to rewrite eq. (\ref{VR}) (upon
replacing $V$, $w$ by $\ve^2v$, $\ve\omega$ in the
continuum limit) in the standard linear plus bilinear form
which reproduces the continuum action $S_{\rm HP}$.

Thus, the continuum action (\ref{S-HP}) can be obtained
from the discrete Regge-type one (\ref{S-VR}) under quite
reasonable assumptions (eqs. (\ref{V-Om-eps}), (\ref{V}),
(\ref{Omega})) defining the continuum limit, and with
partial use of the equations of motion. Since the most
general case when area tensors are independent variables
was considered, the same is valid also for the tetrad
(bivector) form of these variables, i. e. for the usual
general relativity and Regge calculus in the
tetrad-connection form. Besides that, we can restrict the
connection matrices $\Omega$ in the action (\ref{S-VR}) to
have the (anti-)selfdual generators, and this still
presents the Regge calculus action \cite{Kha}. Evidently,
the continuum limit action in this case will be given just
by eq. (\ref{S-HP}) where now $\omega_\lambda^{ab}$ and
therefore $V_{\lambda\mu}^{ab}$ are restricted to be
(anti-)selfdual matrices.

The remaining problem concerning the interrelation between
the continuum and discrete area-connection theories is that
of validity of an analog of the Friedberg-Lee theorem
\cite{FriLee}. That is, whether discrete Regge-type action
can be obtained by the exact calculation of the continuum
action on a particular distribution of the $\omega$, $\pi$
fields, a kind of conical singularities? The positive
answer would mean that area-connection generalisation of
Regge calculus be the second example (after usual Regge
calculus) of the discrete {\it minisuperspace} theory which
at the same time is able to approximate the continuum
counterpart with arbitrarily large accuracy.

\bigskip

The present work was supported in part by the Russian
Foundation for Basic Research through Grant No.
01-02-16898, through Grant No. 00-15-96811 for Leading
Scientific Schools and by the Ministry of Education Grant
No. E00-3.3-148.

\end{document}